\documentclass[preprint,
amsmath,amssymb]{jpsj3}
\usepackage{txfonts}
\usepackage{amsmath,amssymb}
\usepackage{comment}
\usepackage{graphicx}
\usepackage{color}
\usepackage{dcolumn}
\usepackage{bm}
\usepackage{url}
\usepackage{braket}
\usepackage{setspace}
\usepackage{ascmac}
\def\ra{\rangle}

\title{Exact expression of the energy gap at first-order quantum phase transitions of a non-stoquastic Hamiltonian}

\author{Masaki Ohkuwa\thanks{ookuwa.m@stat.phys.titech.ac.jp} and Hidetoshi Nishimori}
\inst{Department of Physics, Tokyo Institute of Technology, Oh-okayama, Meguro-ku, Tokyo 152-5551, Japan} 
\date{\today}%
\abst{We study the energy gap between the ground state and the first excited state of a mean-field-type non-stoquastic Hamiltonian by a semi-classical analysis. The fully connected mean-field model with $p$-body ferromagnetic interactions under a transverse field has a first-order quantum phase transition for $p\ge 3$. This first-order transition is known to be reduced to second order for $p\ge 5$ by an introduction of antiferromagnetic transverse interactions, which makes the Hamiltonian non-stoquastic. This reduction of the order of transition means an exponential speedup of quantum annealing by adiabatic processes because the first-order transition is shown to have an exponentially small energy gap whereas the second order case does not. We apply a semi-classical method to analytically derive the explicit expression of the rate of the exponential decay of the energy gap at first-order transitions. The result reveals how the property of first-order transition changes as a function of the system parameters.  We also derive the exact closed-form expression for the critical point where the first-order transition line disappears within the ferromagnetic phase. These results help us understand how the antiferromagnetic transverse interactions affect the performance of quantum annealing by controlling the effects of non-stoquasticity in the Hamiltonian.
}

\begin{document}
\maketitle
\section{\label{sec:level1}Introduction\protect}
Quantum annealing (QA) is a metaheuristic, a generic approximate algorithm, to solve combinatorial optimization problems by quantum fluctuations \cite{Kadowaki1998,Kadowaki1998b,Brooke1999,Santoro2002,Santoro2006,Das2008,Tanaka_book2017}. Combinatorial optimization is a class of problems to find the global minimum of a cost function of many discrete variables. The time necessary to solve a combinatorial optimization problem grows typically exponentially as a function of the problem size, a hallmark of difficult problems. There exist various methods developed to solve each of those difficult problems, among which simulated annealing (SA) \cite{Kirkpatrick1983} is a classical metaheuristic that uses classical stochastic processes in analogy with thermal fluctuations in statistical mechanics.  QA has often been compared with SA to investigate if quantum effects may ever be useful to accelerate computation \cite{Ronnow2014}, a recent example of which is a careful study showing evidence of a limited quantum speedup \cite{AlbashLidar2017}.

The cost function of a combinatorial optimization problem can be expressed  as the Hamiltonian of an Ising model with complicated interactions \cite{Lucas2014}. This means that the search for an optimal solution is analogous to the identification of the ground state of an Ising spin glass, one of the most difficult problems in statistical physics \cite{Nishimori_book_sg}. Thus, the study of the performance of QA has important significance from the viewpoint not just of computer science but also of physics.

In the initial stage of QA, one applies a strong transverse field to the system to keep the system state in the trivial ground state with all spins (qubits) pointing to the same direction along the transverse field. One then reduces the transverse field toward zero, and the system evolves following the time-dependent  Schr\"odinger equation.  Since the initial trivial state and final non-trivial state are completely different, a quantum phase transition takes place at a certain strength of the transverse field during the course of QA.
Let us consider a process in which the strength of the transverse field, relative to the magnitude of the coefficients in the Ising model, is reduced from a very large value to zero in a finite amount of time $\tau$ \cite{Farhi2000,Farhi2002}.  The system stays close to the instantaneous ground state if $\tau$ is very large compared to the inverse square of the minimum energy gap between the ground state and the first excited state, $\tau\gg\Delta^{-2}_{min}$, according to the adiabatic theorem of quantum mechanics (see e.g. \cite{AlbashLidar2016,Morita2008}). If there exists a second-order quantum phase transition at a certain value of the transverse field, the energy gap decreases polynomially as a function of the system size according to finite-size scaling \cite{Nishimori_book}. Then, the problem can be solved in polynomial time and is considered easy. In contrast, if the quantum phase transition is of first order, the minimum gap decays exponentially as the system size increases, and the computation time increases exponentially. This implies that a first-order transition sets a serious bottleneck for QA. In this way, the study of quantum phase transition in the transverse field Ising model gives important information on the performance of QA.

The infinite-range Ising model with $p$-body ferromagnetic interactions under a transverse field ($p$-spin model) is a prototype problem that has been studied in the context of QA due to its amenability to analytical investigations \cite{Jorg2010,Bapst2012}.  It is known that this model has a first-order quantum phase transition for $p\ge 3$, and is therefore hard to solve by the conventional QA for the reason described above. Nevertheless, it was shown that this first-order transition can be avoided by the introduction of antiferromagnetic transverse interactions into the Hamiltonian, by which first-order phase transitions are reduced to second order \cite{Seki2012}. A similar phenomenon has been shown to exist in the Hopfield model \cite{Seki2015}, which involves randomness in interactions and is thus much more nontrivial compared with the simple $p$-spin model. Since the antiferromagnetic transverse interactions make the Hamiltonian non-stoquastic \cite{Bravyi2008}, the above example of the reduction of the order of phase transition may be taken as a remarkable example in which a non-stoquastic Hamiltonian shows an exponential speedup over its stoquastic counterpart \cite{NishimoriTakada2017}.  See also Refs. \citen{Farhi2012,Crosson2014,Hormozi2017} for numerical evidence that a non-stoquastic Hamiltonian may sometimes accelerate the performance of QA.

The goal of the present paper is to analyze the properties of first-oder quantum phase transitions in the $p$-spin model with antiferromagnetic transverse interactions.  In particular, we use the semi-classical method \cite{Braun1993,Garg1998,Garg1999,Bapst2012,Kechedzhi2016} to derive the explicit formula of the energy gap as a function of system parameters. Similar semi-classical analyses of the energy gap have already been reported in the context of QA for stoquastic Hamiltonians \cite{Bapst2012,Isakov2016,Jiang2017,Kechedzhi2016}. Our contribution is to generalize these results to the non-stoquastic case, by which we show how the energy gap behaves as the effects of the antiferromagnetic transverse interactions become prominent and quantum fluctuations play important roles \cite{Susa2017}.

We formulate and solve the problem in the next section. The final section concludes the paper.

\section{Semi-classical analysis of the phase diagram and the energy gap}

We first define the problem and then proceed to describe a semi-classical  approach to analyze the phase diagram and the energy gap.

\subsection{Problem formulation}

The problem we study in this paper is described by the Hamiltonian
\begin{align}
\label{eq:hamiltonian1}
\hat{H}(s,\lambda)=s\Big(\lambda\hat{H}_0+(1-\lambda)\hat{V}_{{\rm TI}}\Big)+(1-s)\hat{V}_{{\rm TF}}\quad (0\le s, \lambda \le 1),
\end{align}
where $H_0$ is the $p$-spin model, $\hat{V}_{{\rm TI}}$ is the antiferromagnetic transverse interaction, and $\hat{V}_{{\rm TF}}$ denotes the usual transverse field,
\begin{align}
\hat{H}_0=-N\left(\frac{1}{N}\sum_{i=1}^N\hat{\sigma}_i^z\right)^p,\quad
\hat{V}_{{\rm TI}}=N\left(\frac{1}{N}\sum_{i=1}^N\hat{\sigma}_i^x\right)^2,\quad
\hat{V}_{{\rm TF}}=-\sum_{i=1}^N\hat{\sigma}_i^x.
\end{align}
The initial values of the parameters are $s=0$ and $\lambda$ arbitrary since $\lambda$ disappears when $s=0$. We then let the system evolve adiabatically toward the goal of $s=\lambda =1$, where only the problem Hamiltonian $\hat{H}_0$ remains.  Notice that the Hamiltonian is stoquastic only when $\lambda=1$.  For any $\lambda$ less than 1, i.e. with the antiferromagnetic transverse interactions $\hat{V}_{{\rm TF}}$, the Hamiltonian is non-stoquastic, and the effects of the term $\hat{V}_{{\rm TI}}$ are stronger for smaller $\lambda$.

When $p>2$ and $\lambda=1$, this model is known to go under a first-order transition as a function of $s$ \cite{Jorg2010}. This first-order transition has been shown to be avoided by the introduction of $\hat{V}_{{\rm TI}}$ \cite{Seki2012}. We choose $p$ to be odd to avoid the trivial double degeneracy for even $p$.

It is convenient to rewrite the Hamiltonian as
\begin{align}
\hat{H}=-s\lambda N\left(\frac{2}{N}\hat{S}_z\right)^p+s(1-\lambda)N\left(\frac{2}{N}\hat{S}_x\right)^2-2(1-s)\hat{S}_x.
\end{align}
where
\begin{align}
\hat{S}_{x}=\frac{1}{2}\sum_{i=1}^N\hat{\sigma}_i^x,\;\;\hat{S}_{y}=\frac{1}{2}\sum_{i=1}^N\hat{\sigma}_i^y,\;\;\hat{S}_{z}=\frac{1}{2}\sum_{i=1}^N\hat{\sigma}_i^z.
\end{align}
The total spin operator
\begin{eqnarray}
\hat{S}^2=\hat{S}_x^2+\hat{S}_y^2+\hat{S}_z^2
\end{eqnarray}
commutes with the Hamiltonian, and therefore the total spin $S$ is conserved in the process of time evolution. In the initial ground state, the eigenvalue of $\hat{S}_x$ is maximum, $S=N/2$. The final ground state belongs to the same subspace. Therefore, we only need to consider the subspace with $S=N/2$. Moreover, it is known that the first excited state state also belongs to the same subspace \cite{Bapst2012}.

For later convenience, we normalize spin operators as $\hat{m}_i=\hat{S}_i/S$. The Hamiltonian per spin is
\begin{eqnarray}
\hat{E}=\frac{\hat{H}}{N}=-s\lambda\hat{m}_z^p+s(1-\lambda)\hat{m}_x^2-(1-s)\hat{m}_x.
\label{operator_E}
\end{eqnarray}
Let us write the eigenstate of the normalized spin operator as
\begin{eqnarray}
\hat{m}_z\ket{S,m}=m\ket{S,m},\;\;\left(m=-1,-1+\frac{1}{S},...,1\right).
\end{eqnarray}
We express the eigenstate of $\hat{E}$ as
\begin{eqnarray}
\ket{\Psi}=\sum_mC_m\ket{S,m}.
\end{eqnarray}
The Schr\"odinger equation can be written in terms of the coefficients as
\begin{align}
\nonumber
&EC_m=\Big(-s\lambda m^p+\frac{s(1-\lambda)}{2}(1-m^2+S^{-1})\Big)C_m\\ 
\nonumber
&-\frac{1-s}{2}\left(\sqrt{(1+m)(1-m+S^{-1})}\,C_{m-1/S}+\sqrt{(1-m)(1+m+S^{-1})}\,C_{m+1/S}\right) \\
\nonumber
&+\frac{s(1-\lambda)}{4}\Big(\sqrt{\left(1-m+2S^{-1}\right)\left(1+m-S^{-1}
\right)\left(1-m+S^{-1}\right)(1+m)}\,C_{m-2/S} \\
&+\sqrt{\left(1+m+2S^{-1}\right)\left(1-m-S^{-1}\right)\left(1+m+S^{-1}
\right)(1-m)}\,C_{m+2/S}\Big).
\end{align}
Since we are interested in the semi-classical limit ($S\gg 1$), we extract the leading terms as
\begin{align}
\label{eq:syure1}
\nonumber
EC_m&=\Big(-s\lambda m^p+\frac{s(1-\lambda)}{2}(1-m^2)\Big)C_m\\ 
\nonumber
&-\frac{1-s}{2}\sqrt{1-m^2}\left(C_{m-1/S}+C_{m+1/S}\right) \\
&+\frac{s(1-\lambda)}{4}(1-m^2)\left(C_{m-2/S} +C_{m+2/S}\right).
\end{align}

\subsection{Classical potential and the phase diagram}

The asymptotic behavior of the present model for $S\gg 1$ can be analyzed by the semi-classical method \cite{Braun1993,Garg1998,Garg1999,Bapst2012,Kechedzhi2016,Isakov2016,Jiang2017}, i.e. an asymptotic expansion in terms of $1/S$. 
Let us write the coefficient $C_m$ as
\begin{eqnarray}
C_m \propto \exp \big(i\Phi(m)\big),
\label{Cm}
\end{eqnarray}
and expand the exponent as
\begin{eqnarray}
\Phi(m)=S\Phi_0(m)+\Phi_1(m)+\mathcal{O}(S^{-1}).
\label{Phim}
\end{eqnarray}
Equation (\ref{eq:syure1}) is then solved to leading order as
\begin{eqnarray}
\label{eq:ekinji}
E=w(m)+t_1(m)\cos\dot{\Phi}_0+t_2(m)\cos(2\dot{\Phi}_0),
\end{eqnarray} 
where
\begin{align}
w(m)&=-s\lambda m^p+\frac{s(1-\lambda)}{2}(1-m^2),\\
\label{eq:t1t2}
t_1(m)&=-(1-s)\sqrt{1-m^2},\;\;
t_2(m)=\frac{s(1-\lambda)}{2}(1-m^2).
\end{align}

The starting point of the semi-classical analysis is to identify the effective classical potential, from which the quantum tunneling rate and the energy gap are calculated.  To this end, we first notice that $\dot{\Phi}_0(m)$ in Eq. (\ref{eq:ekinji}) can be identified with the momentum. The operator $\hat{q}$ defined by
\begin{equation}
\hat{q}=-i\frac{d}{dm}
\end{equation}
shifts the argument of $C(m)$ as
\begin{equation}
C(m+ j/S)=e^{ij\hat{q}/S}\, C(m)~(j=\pm 1, \pm 2,\cdots).
\end{equation}
Then, Eq. (\ref{eq:syure1}) can  be written as
\begin{align}
EC(m)=w(m)C(m)+t_1(m) \cos \left(\frac{\hat{q}}{S}\right)\, C(m)
+t_2(m) \cos \left(\frac{2\hat{q}}{S}\right) \, C(m).
\label{ECm0}
\end{align}
Comparison of this equation with Eq. (\ref{eq:ekinji}) reveals that $S\dot{\Phi}_0$ can be identified with the momentum in the classical limit. This means that the classical motion is allowed in the range where $\dot{\Phi}_0$ is real and thus $\cos\dot{\Phi}_0$ lies between $-1$ and 1.

\begin{figure*}[thb]
	\centering
		\includegraphics[width=6cm,clip]{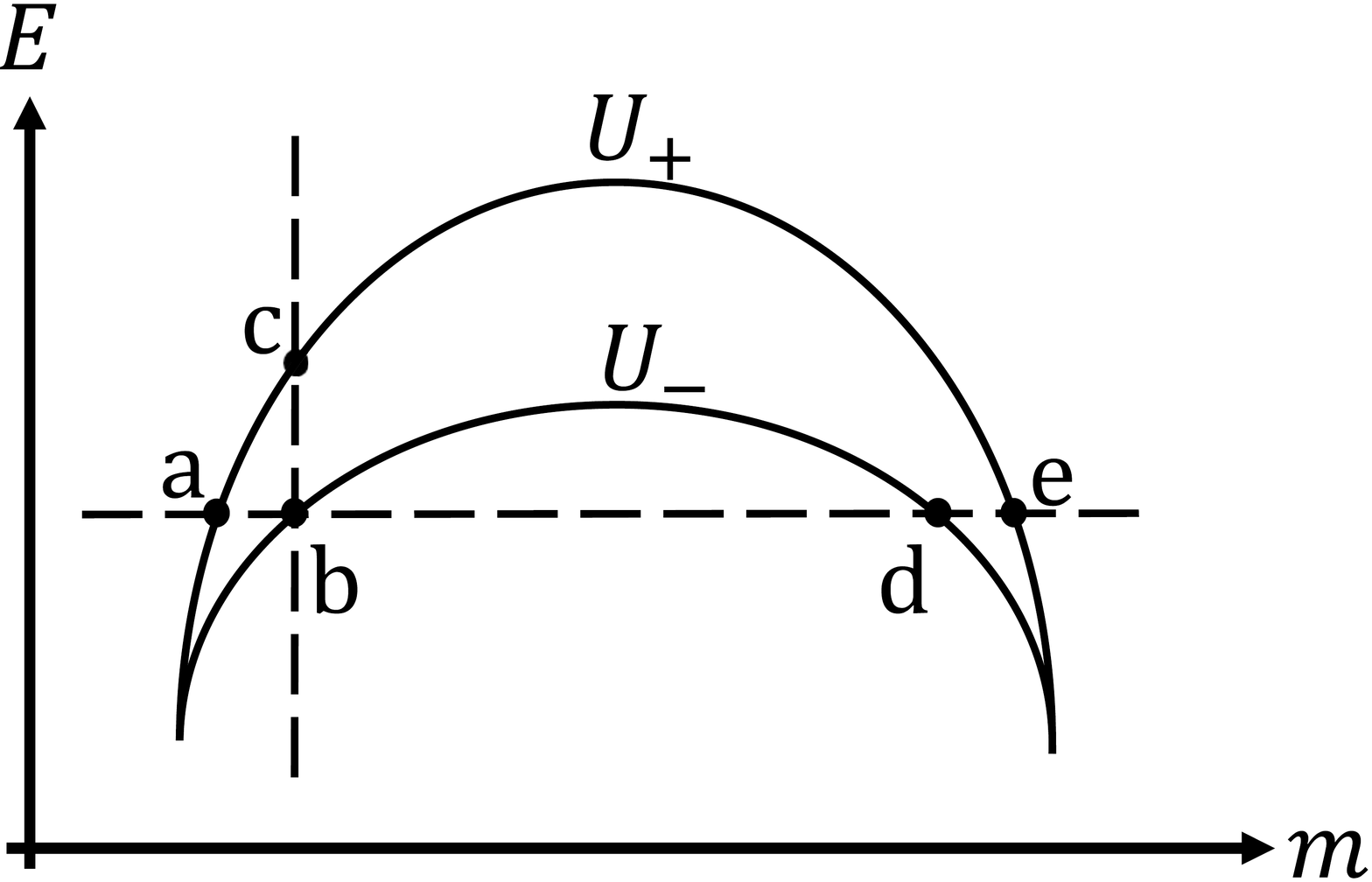}
	\caption{For a fixed value of $m$, the classical energy can take values between b and c.  This fact is translated into the range between a and b for a fixed value of $E$.  Also the range between d and e is allowed classically}
\label{fig:potential}
\end{figure*}

Let us write Eq. (\ref{eq:ekinji}) as
\begin{align}
E=w(m)+2t_2(m)\left(\cos\dot{\Phi}_0(m)+\frac{t_1(m)}{4t_2(m)}\right)^2
-\frac{t_1^2(m)}{8t_2(m)}-t_2(m),
\label{eq:eheihou}
\end{align}
where
\begin{equation}
-\frac{t_1(m)}{4t_2(m)}=\frac{1-s}{2s(1-\lambda)\sqrt{1-m^2}}~(\ge 0).
\label{eq:extremum}
\end{equation}
The classical potential is derived as follows.  For a given fixed value of $m$, the energy $E$ can assume values corresponding to real values of $\dot{\Phi}_0$, the latter condition coming from  real values of the momentum for a classical trajectory.  As depicted in Fig. \ref{fig:potential}, we define the upper bound of the energy as $U_{+}$ (point c in Fig. \ref{fig:potential}) and the lower bound as $U_{-}$ (point b) at a given $m$ as obtained by changing $\cos\dot{\Phi}_0$ in the range $[-1,1]$ for fixed $m$. The energy can take values between these limits.  If we translate this fact into the classically allowed range of $m$ for a given $E$, the parameter $m$ can take values between  $U_{+}$ and  $U_{-}$ for fixed $E$, e.g. between points a and b as well as between d and e in Fig. \ref{fig:potential}.  The region between b and d is forbidden classically since the momentum becomes imaginary.
To find an explicit expression of the classical potential, we therefore have to minimize Eq. (\ref{eq:eheihou}) with respect to $\cos\dot{\Phi}_0$ in the range between $-1$ and 1.  It is necessary to consider the cases with $1-s>2s(1-\lambda)$ and $1-s<2s(1-\lambda)$ separately since the right hand side of Eq. (\ref{eq:extremum}) may or may not exceed 1 depending on the direction of the inequality. Before going further, we notice that the condition
\begin{eqnarray}
\label{eq:secondline}
1-s=2s(1-\lambda)
\end{eqnarray}
is known to give the second-order transition line in the $s$-$\lambda$ phase diagram \cite{Seki2012}. The second order phase transition thus corresponds to the border of classification of the classical potential.

When $1-s>2s(1-\lambda)$, the energy $E$ assumes its minimum at $\cos\dot{\Phi}_0=1$ according to Eqs. (\ref{eq:eheihou}) and (\ref{eq:extremum}). Then the classical potential is
\begin{align}
\nonumber
U_1(m)&:=w(m)+t_1(m)+t_2(m)\\
&=-s\lambda m^p-(1-s)\sqrt{1-m^2}+s(1-\lambda)(1-m^2).\label{potential1}
\end{align}
This coincides with the classical potential naively obtained by the replacement of $\hat{m}_z$ by $m$ and $\hat{m}_x$ by $\sqrt{1-m^2}$ in Eq. (\ref{operator_E}) as was done in Refs. \citen{Seoane2012,NishimoriTakada2017}.

The situation is slightly more complicated when $1-s<2s(1-\lambda)$. Suppose that the following relation holds,
\begin{eqnarray}
1\le-\frac{t_1(m)}{4t_2(m)}=\frac{1-s}{2s(1-\lambda)\sin\theta},\label{cond1}
\end{eqnarray}
which is rewritten in terms of $m(=\cos\theta)$ as
\begin{eqnarray}
|m|\ge\sqrt{1-\left(\frac{1-s}{2s(1-\lambda)}\right)^2}=: m_0
\end{eqnarray}
If $|m|>m_0$, i.e. if Eq. (\ref{cond1}) holds, the condition $\cos\dot{\Phi}_0(m)=1$ gives the minimum of the energy $E$ of Eq. (\ref{eq:eheihou}) as a function of $\cos\dot{\Phi}_0(m)$.  The resulting expression is the same $U_1(m)$ of Eq. (\ref{potential1}) as in the case of $1-s>2s(1-\lambda)$. On the other hand, if $|m|<m_0$, the minimum of $E$ is at $\cos\dot{\Phi}_0(m)=-t_1/4t_2$. Therefore the explicit expression of the classical potential for $1-s<2s(1-\lambda)$ is
\begin{align}
U(m)=\begin{cases}
	w(m)+t_1(m)+t_2(m)=U_1(m)  & (|m|\ge m_0) \\
	w(m)-\displaystyle\frac{t_1^2(m)}{8t_2(m)}-t_2(m)=:U_2(m) & (|m|\le m_0),
  \end{cases}
\end{align}
where $U_2(m)$ is expressed explicitly as
\begin{eqnarray}
\label{eq:U2}
U_2(m)=-s\lambda m^p-\frac{(1-s)^2}{4s(1-\lambda)}.
\end{eqnarray}
It is clear that the minimum of $U(m)$ as a function of $m$ is located in the range $|m|\ge m_0$ since $U_2(m)$ is a monotonically decreasing function of $m$. This means that we may consider only $U_1(m)$ as long as we are interested in the minimum of $U(m)$ as a function of $m$ to discuss the ground state. This observation justifies the naive classical analysis using Eq. (\ref{potential1}) to draw the phase diagram in Refs. \citen{Seoane2012,NishimoriTakada2017}.

\begin{figure*}[thb]
\centering
    \begin{tabular}{c}
      \begin{minipage}{0.33\linewidth}
			\centering
          \includegraphics[width=5cm,clip]{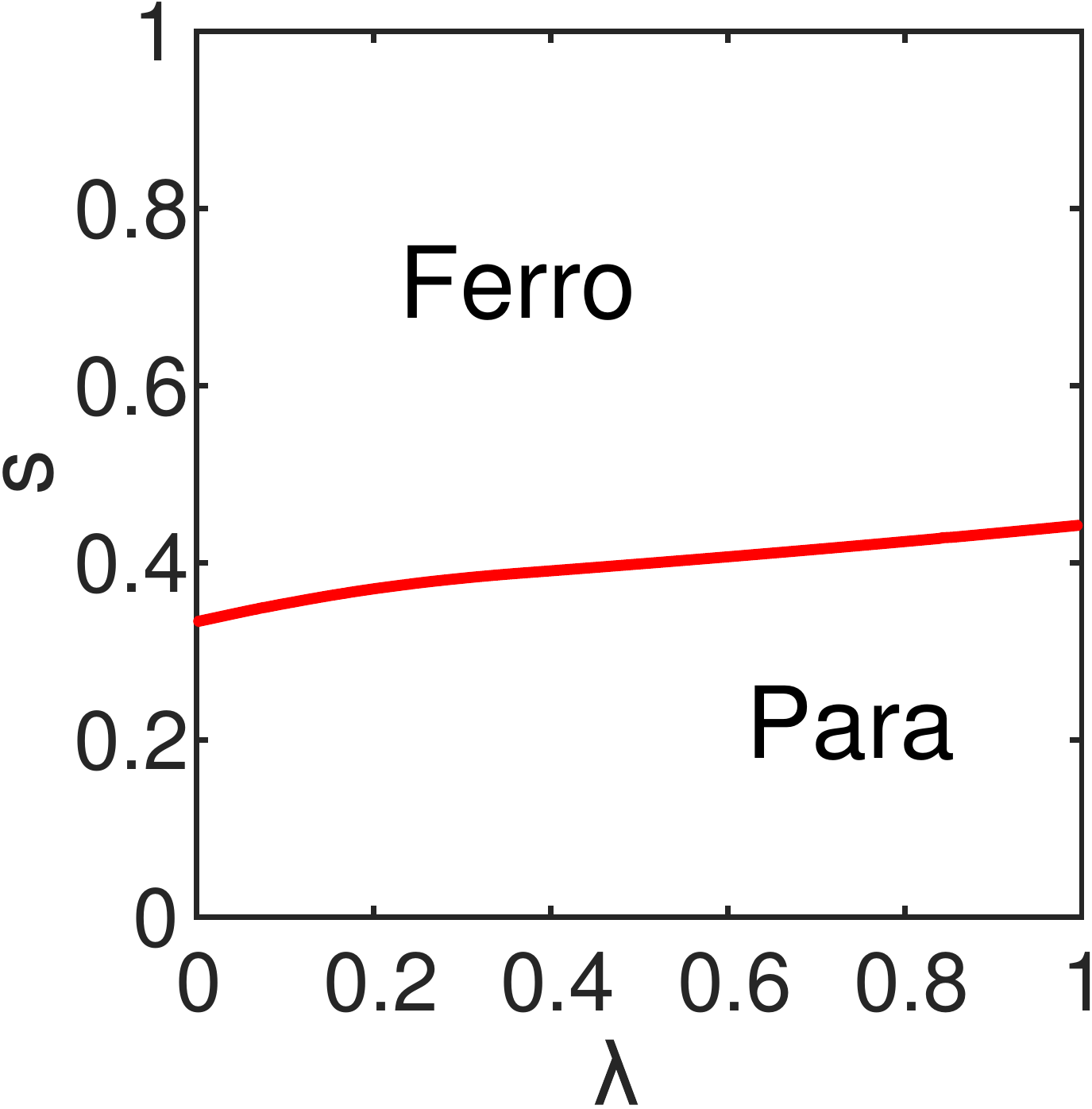}
          \hspace{1.6cm} $p=3$
      \end{minipage}
      \begin{minipage}{0.33\linewidth}
        \begin{center}
          \includegraphics[width=5cm,clip]{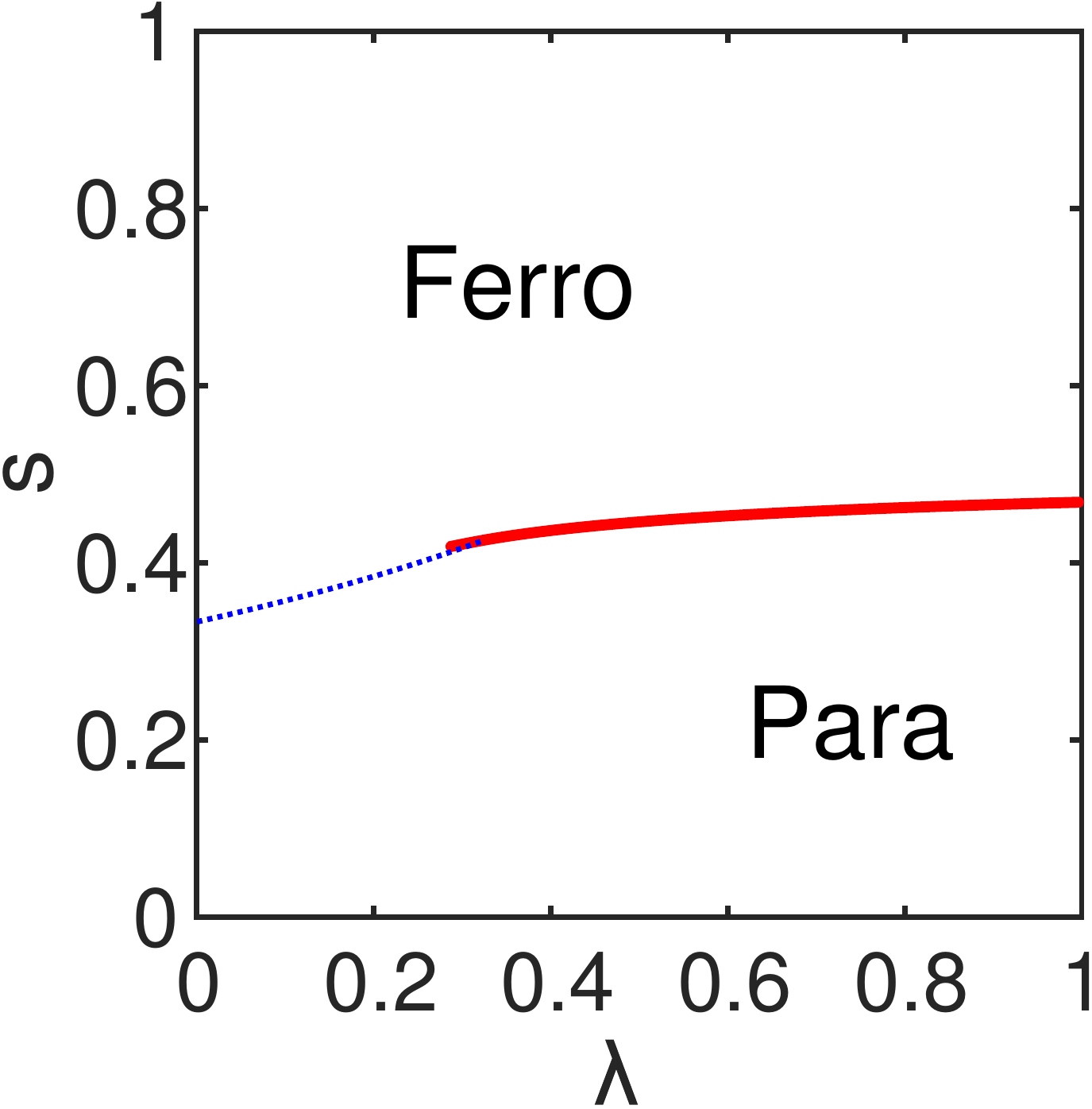}
          \hspace{1.6cm} $p=5$
        \end{center}
      \end{minipage}
      \begin{minipage}{0.33\linewidth}
        \begin{center}
          \includegraphics[width=5cm,clip]{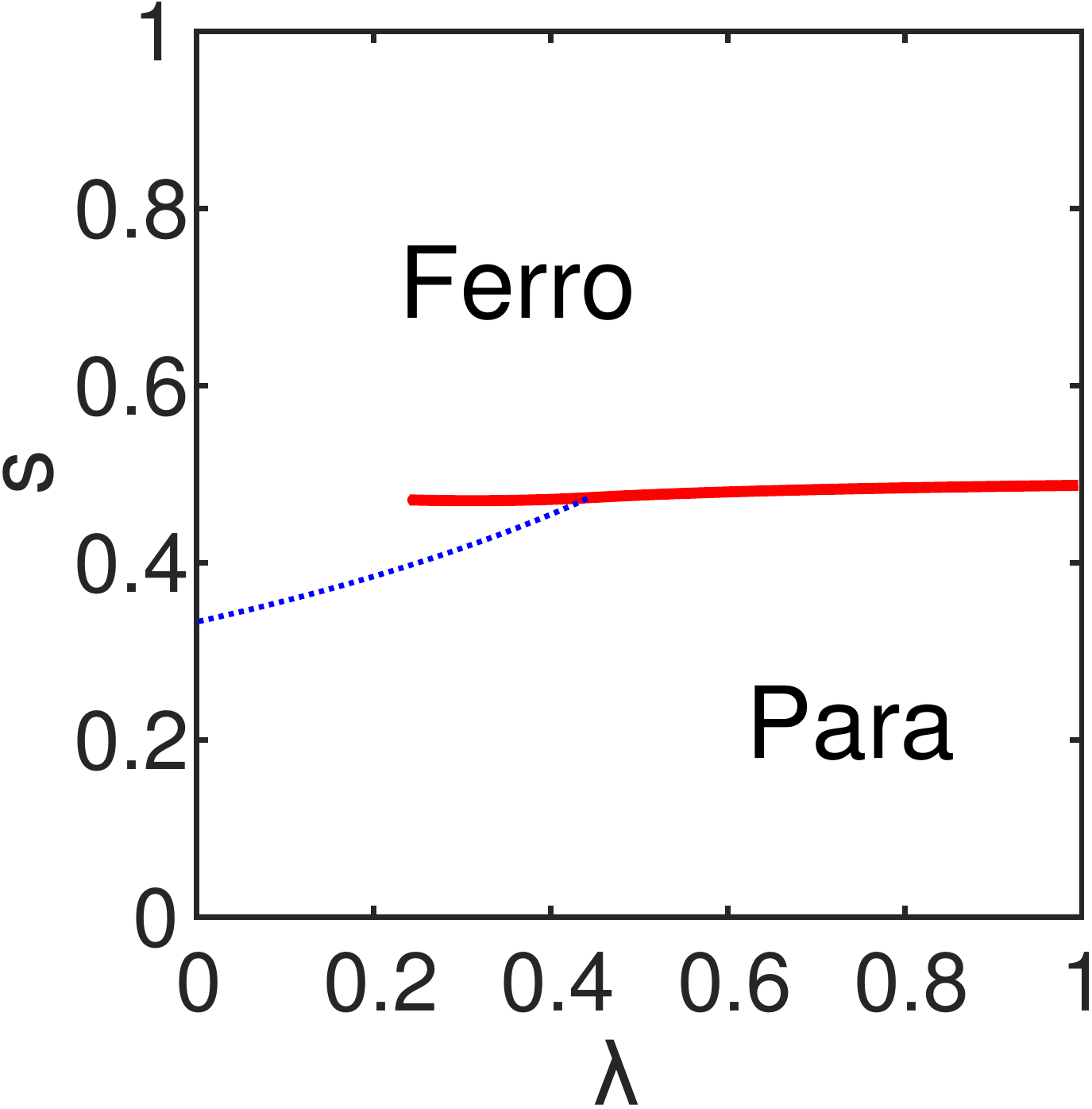}
          \hspace{1.6cm} $p=11$
        \end{center}
      \end{minipage}
    \end{tabular}
    \caption{Phase diagram on the $s$-$\lambda$ plane for $p=3$ (left), $p=5$ (middle), and $p=11$ (right). The red bold curve is a line of first-order transitions between the quantum paramagnetic phase and the ferromagnetic phase. The blue dotted curve is a line of second order transitions. 
}
\label{fig:phasediagrams}
\end{figure*}

Analyses of the potential $U_1(m)$ to draw the phase diagram have already been carried out in detail in Ref. \citen{Seoane2012}, and we just quote the result. Examples of the phase diagram are shown in Fig. \ref{fig:phasediagrams} for $p=3, 5$ and 11. The latter two cases have a line of second-order transitions for smaller $\lambda$ that replaces the first-order line for larger $\lambda$.  This means that we can avoid an exponentially small gap at a first-order transition by choosing an appropriate path in the phase diagram from $(s=0, \lambda)$ to $s=\lambda=1$.

\subsection{Exact location of the critical point}

It is possible to derive a closed-form expression of the critical point where the line of first-order phase transitions terminates within the ferromagnetic phase as in Fig. \ref{fig:phasediagrams}, seen most clearly for $p=11$. The Landau theory of phase transitions tells us that the derivatives of the free energy vanish up to third order at such a critical point \cite{Nishimori_book}. Let us denote the magnetization at the critical point as $m_*=\cos\theta_*$.  Then, the critical condition derived from the derivatives of the potential $U_1(\cos\theta)$ is
\begin{align}
\label{eq:firstder}
&ps\lambda\cos^{p-2}\theta_*\sin\theta_*-(1-s)+2(1-\lambda)s\sin\theta_*=0\\
\label{eq:secondder}
&ps\lambda\cos^{p-2}\theta_*\left(1-(p-2)\frac{\sin^2\theta_*}{\cos^2\theta_*}\right)+2s(1-\lambda)=0
\\
\label{eq:thirdder}
&p-7-(p-1)\cos2\theta_*=0.
\end{align}
Equation (\ref{eq:thirdder}) is solved as
\begin{equation}
m_*=\sqrt{\frac{p-4}{p-1}},
\end{equation}
which is the value of the magnetization at the critical point.
Equations (\ref{eq:firstder}) and (\ref{eq:secondder}) then give the critical values of $\lambda$ and $s$ as
\begin{align}
\label{eq:lambdacritical}
\lambda_*&=\frac{1}{1+pm_*^{p-4}}\\
\label{eq:scritical}
s_*&=\frac{1+pm_*^{p-4}}{1+pm_*^{p-4}\left(1+\sqrt{\frac{3}{p-1}}\left(m_*^2+2\right)\right)}.
\end{align}
These equations give the exact location of the critical point, which we have confirmed to agree with the numerical result for several values of $p$.

\subsection{Closed-form expression of the energy gap}

The energy gap at a first-order phase transition is given as the overlap of two wave functions $|\Psi_1\ra$ and $|\Psi_2\ra$ localized at two degenerate minima, $m=m_1$ and $m=m_2$, of the classical potential $U_1(m)$, according to the general semi-classical argument as described in Ref.  \citen{Bapst2012}. To calculate the overlap, we need an explicit formula for the wave functions $\phi_1(m)=\langle S, m|\Psi_1\ra$ and $\phi_2(m)=\langle S, m|\Psi_2\ra$.  Since $m_1$ and $m_2$ are the two degenerate minima of the classical potential at a first-order transition, the wave functions $\phi_1(m)$ and $\phi_2(m)$ should decay exponentially away from those minima. Also, the momentum $\dot{\Phi}_0$ should be imaginary for quantum tunneling to take place, and so we write $\Phi_0(m)=i\Psi (m)$ with real $\Psi (m)$. We therefore find the following expression of the wave function from Eqs. (\ref{Cm}) and (\ref{Phim}),
\begin{equation}
\phi_j(m)= \exp \left(-S \left|\int_{m_j}^m \dot{\Psi} (m')dm'\right|\right)\quad (j=1, 2).
\end{equation}
The coefficient $\alpha$ of the exponential decay of the gap $e^{-\alpha N}$ is then given in terms of the overlap as
\begin{align}
\alpha &= -\lim_{N\to\infty}\frac{1}{N} \ln \langle \Phi_1|\Phi_2\ra \nonumber\\
&=-\lim_{N\to\infty}\frac{1}{N}\ln \int_{-1}^1 \phi_1(m) \phi_2(m)\, dm \nonumber\\
&=-\lim_{S\to\infty}\frac{1}{2S}\ln \int_{-1}^1 \exp \left(-S\left|\int_{m_1}^m \dot{\Psi}(m')dm'\right|
-S\left|\int_{m_2}^m \dot{\Psi}(m')dm'\right|\right)\, dm \nonumber\\
&=-\frac{1}{2}\sup_m \left\{- \left|\int_{m_1}^m \dot{\Psi}(m')dm'\right|- \left|\int_{m_2}^m \dot{\Psi}(m')dm'\right| \right\}.
\label{alpha_sup}
\end{align}
The function $\dot{\Psi}(m)$ is defined by Eq. (\ref{eq:ekinji}) with $\dot{\Phi}_0(m) =i\dot{\Psi}(m)$, which changes cosine to hyperbolic cosine in Eq. (\ref{eq:ekinji}). The inverse of hyperbolic cosine has two branches, and we choose the positive one, $\dot{\Psi}(m)\ge 0$. Then the supremum in Eq. (\ref{alpha_sup}) is achieved  when $m$ lies in the range $m_1<m<m_2$, where we assumed $m_1<m_2$. We therefore have an $m$-independent argument of supremum  \cite{Bapst2012},
\begin{equation}
-\int_{m_1}^m \dot{\Psi}(m')dm'-\int_{m}^{m_2} \dot{\Psi} (m')dm' =-\int_{m_1}^{m_2}\dot{\Psi}(m')dm'.
\end{equation}
Other possibilities of $m$ in Eq. (\ref{alpha_sup}) lead to a negative term additional to the above expression such as
\begin{equation}
-2\int_{m_2}^m \dot{\Psi}(m')dm'
\end{equation}
for $m>m_2$ and
\begin{equation}
-2\int_{m}^{m_1} \dot{\Psi}(m')dm'
\end{equation}
for $m<m_1$. Therefore the final expression of the coefficient is
\begin{align}
\alpha&=\frac{1}{2}\int^{m_2}_{m_1}\dot{\Psi}(m')dm',
\nonumber\\
\dot{\Psi}(m)&={\rm arccosh}\,\left(\frac{-t_1(m) -\sqrt{t_1^2(m)-8t_2(m)\left(w(m)-t_2(m)-E\right)}}{4t_2(m)}\right)\nonumber\\
&={\rm arccosh}\left(\frac{(1-s)-\sqrt{(1-s)^2-4s(1-\lambda)(-s\lambda m^p-E)}}{2s(1-\lambda)\sqrt{1-m^2}}\right).
\label{eq:arccosh}
\end{align}
Here, $E$ is the classical energy at the transition point,
\begin{eqnarray}
E=U_1(m_1)=U_1(m_2).
\end{eqnarray}
Equation (\ref{eq:arccosh}) for $\dot{\Psi}(m)$ has been derived from Eq. (\ref{eq:ekinji}) under the replacement $\dot{\Phi}_0=i\dot{\Psi}$, which changes $\cos (x)$ to $\cosh (x)$.  Physically, this means that quantum tunneling takes place in the classically forbidden range of the parameter $m_1<m<m_2$ (between points b and d in Fig. \ref{fig:potential}).

\begin{figure*}[h]
	\centering
		\begin{tabular}{cc}
			\begin{minipage}[t]{0.5\hsize}
				\centering
				\includegraphics[width=7cm]{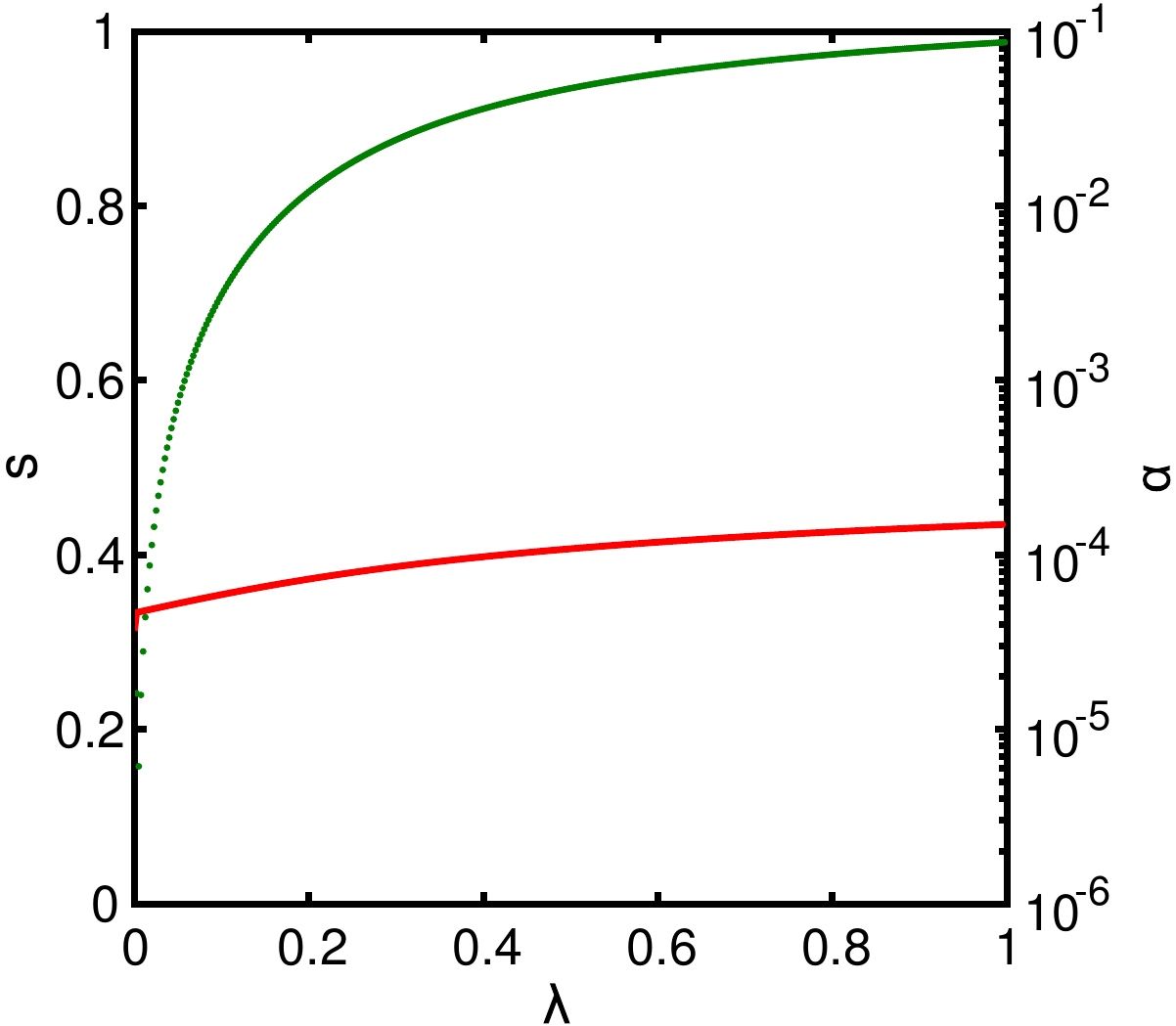}
				\hspace{1.6cm} $p=3$
			\end{minipage}
			\begin{minipage}[t]{0.5\hsize}
				\centering
				\includegraphics[width=7cm]{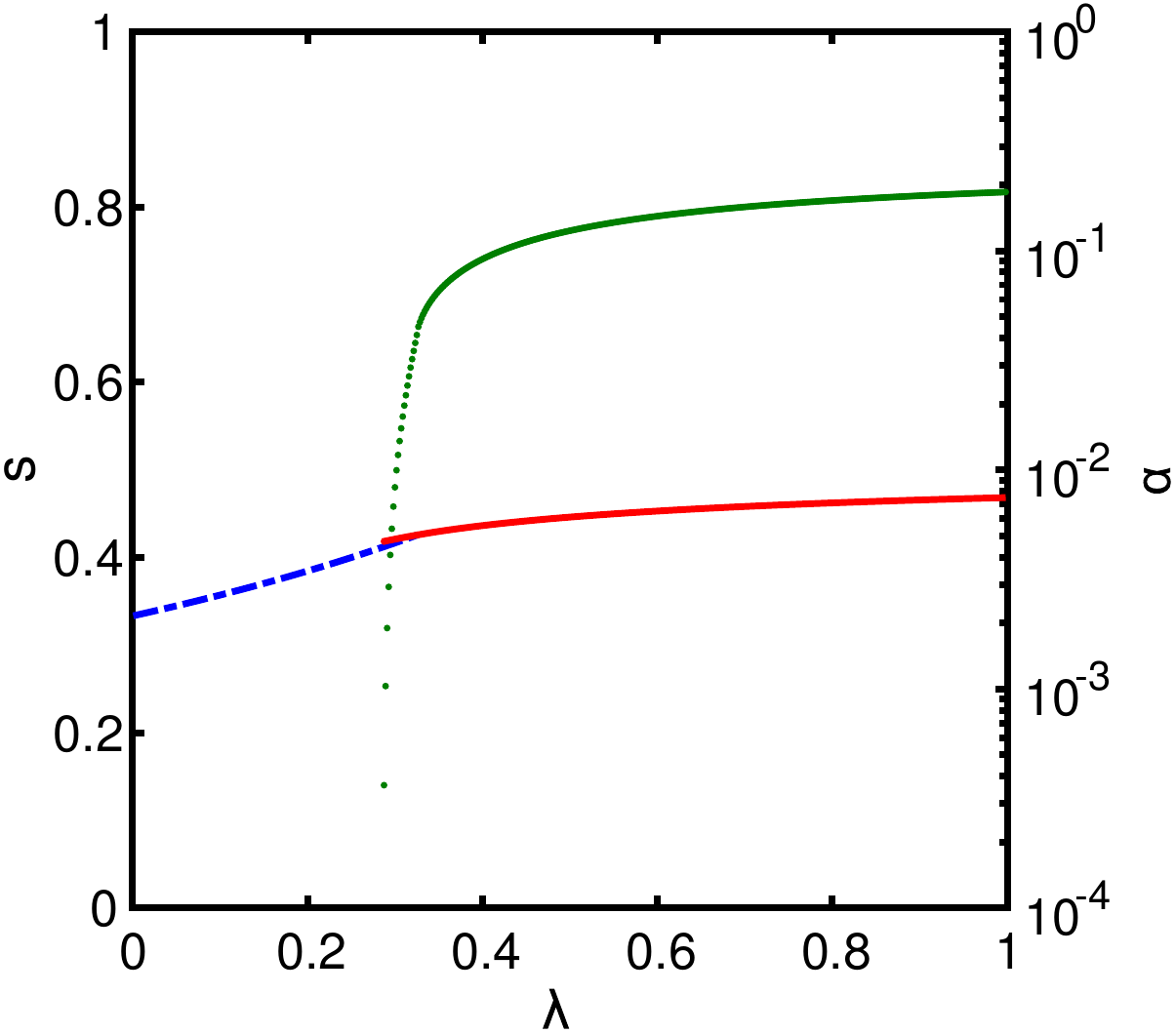}
				\hspace{1.6cm} $p=5$
			\end{minipage}
\\
			\begin{minipage}[c]{0.5\hsize}
				\centering
				\includegraphics[width=7cm]{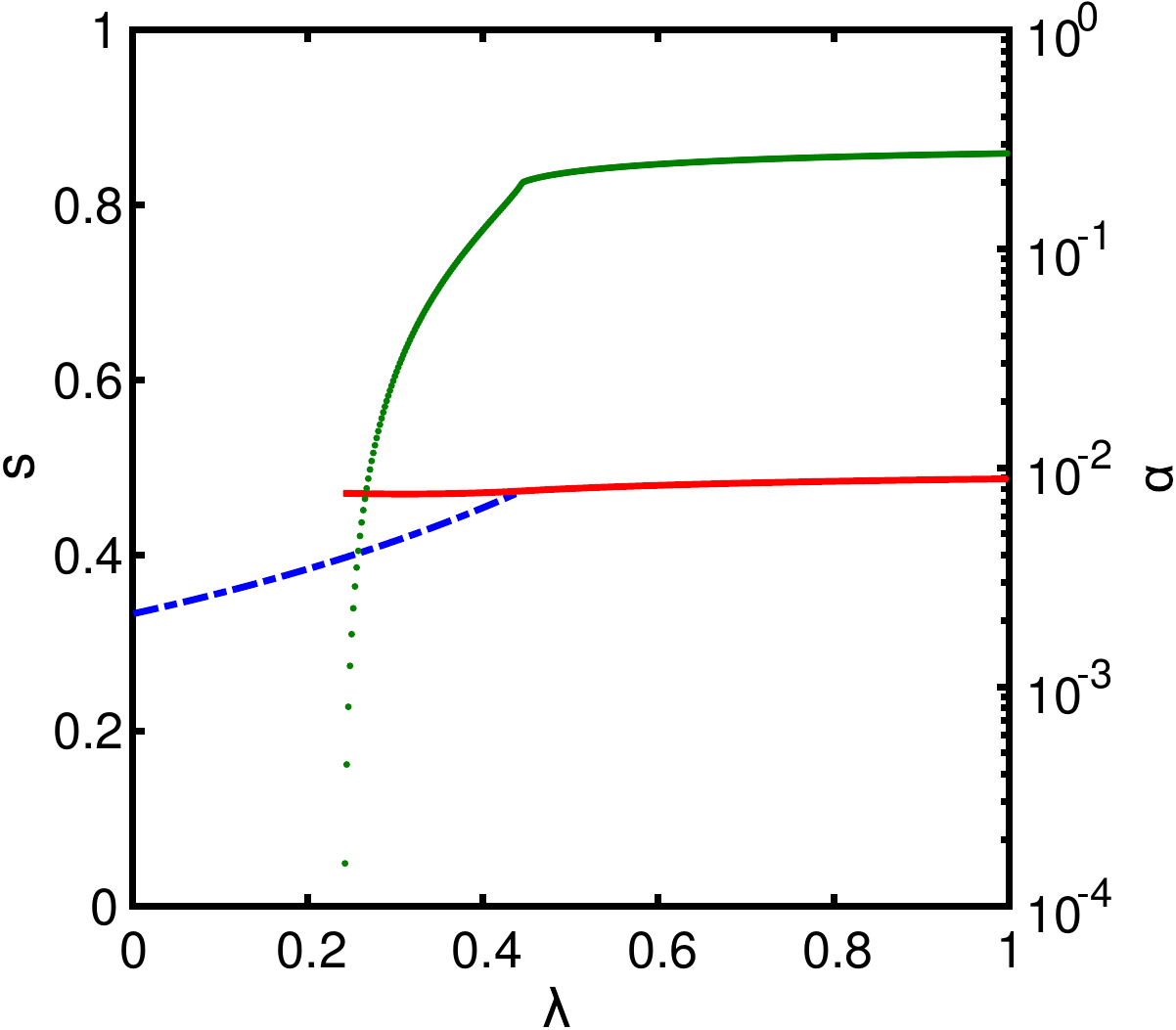}
				\hspace{1.6cm} $p=11$		
			\end{minipage}
			\begin{minipage}[c]{0.5\hsize}
				\centering
				\includegraphics[width=7cm]{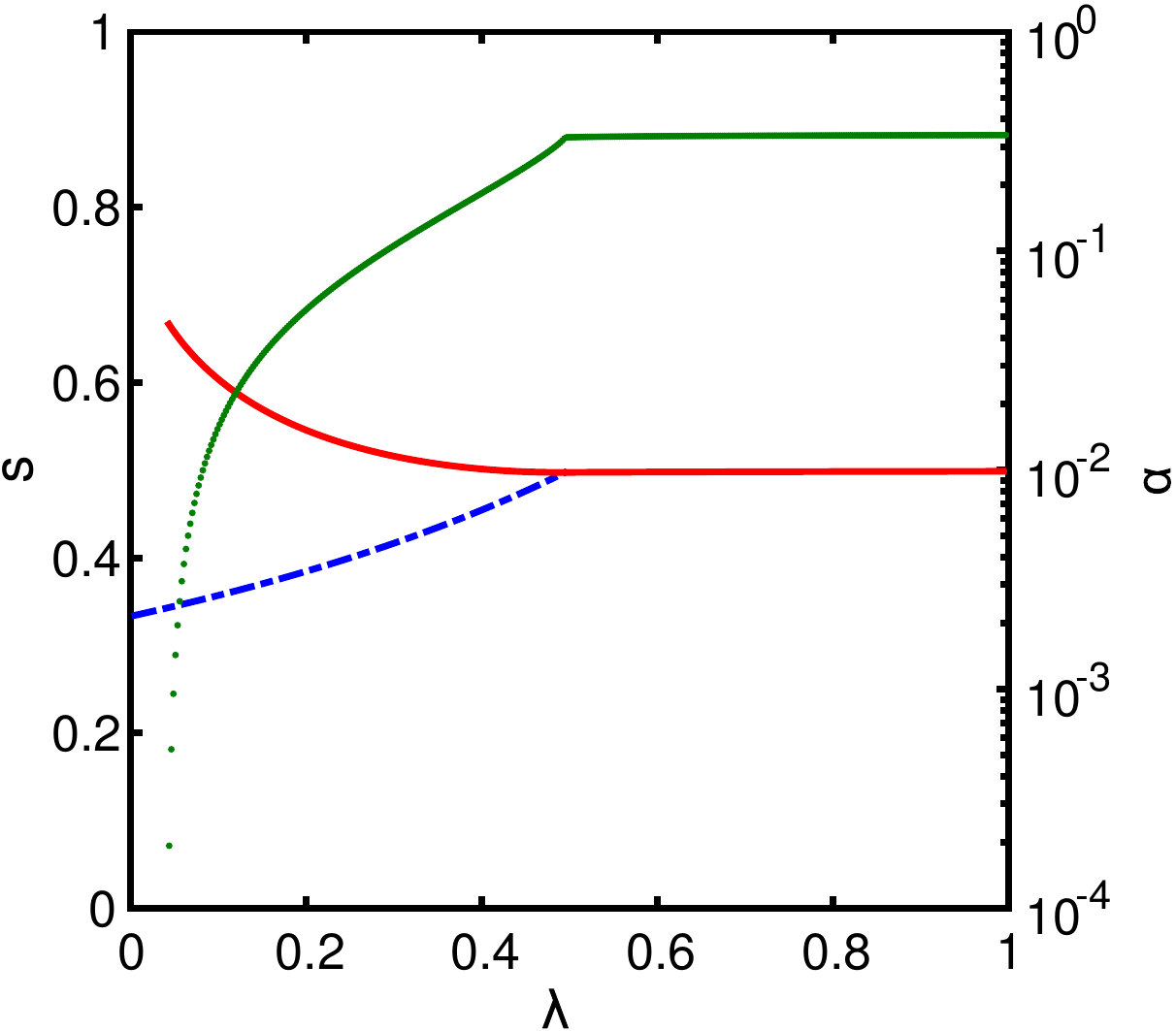}
				\hspace{1.6cm} $p=101$
			\end{minipage}
		\end{tabular}
\caption{The coefficient $\alpha$ of the energy gap closing rate along the first phase transition line on the $s$-$\lambda$ phase diagram. The horizontal axis represents $\lambda$, the right vertical axis is for $\alpha$ in a logarithmic scale, the left axis is $s$. The red bold line represents first-order phase transitions. The blue dash-dotted line denotes second-order phase transitions, $s=1/(3-2\lambda)$. The green dots are for $\alpha$.}
\label{fig:alpha-ABCD}
\end{figure*}

Values of $\alpha$ are plotted in Fig. \ref{fig:alpha-ABCD} fro $p=3, 5, 11,$ and 101.  Except for $p=3$, it is seen that $\alpha$ decreases mildly as $\lambda$ decreases from 1, until the behavior changes at the point where the first-order line meets the second-order line. Beyond this point, $\alpha$ decreases very rapidly toward 0. The dependence of $\alpha$ on $\lambda (>1/2)$ is weak for $p$ large and is close to 1/2, which is the value of $\alpha$ at $\lambda=1$ in the limit $p\to\infty$ \cite{Bapst2012}. The case of $p=3$ is an exception since the transition is always of first order for $0<\lambda \le 1$ \cite{NishimoriTakada2017}.

\section{Conclusion}

We have derived a closed-form expression of the energy gap at first-order quantum phase transitions in the $p$-spin model with antiferromagnetic transverse interactions. This is the first case, as far as we are aware of, that such an expression has been obtained for a non-stoquastic Hamiltonian.  The result reveals that the rate of gap closing depends on the system parameters $s$ and $\lambda$ relatively mildly until the first-order transition line hits the second-order line as the parameter $\lambda$ decreases to about 0.5.  After the two transition lines meet, the coefficient $\alpha$ becomes rapidly decreasing as $\lambda$ further decreases and eventually vanishes at a critical point within the ferromagnetic phase. These behaviors have been anticipated by direct numerical diagonalization of the Hamiltonian \cite{Seki2012}, and our contribution is to have derived the explicit analytical formula, which may lay a foundation for further developments.  The exact location of the critical point within the ferromagnetic phase has also been derived. These results can be taken as the final step to clarify the static properties of the non-stoquastic $p$-spin model from the perspective of quantum annealing.

It would be an interesting future direction to take into account finite temperatures to understand thermal effects on the performance of QA, possibly following the line layed out in Ref. \citen{Kechedzhi2016}.

\begin{acknowledgments}
This work was supported by the JSPS KAKENHI Grant No. 26287086.
\end{acknowledgments}

\bibliographystyle{jpsj.bst}

\end{document}